# DESIGNING AND IMPLEMENTING E-JUSTICE SYSTEMS: AN INFORMATION SYSTEMS APPROACH TO REGIONAL TRIAL COURT CASE DOCKET MANAGEMENT IN NORTHERN MINDANAO, PHILIPPINES

**Benzar Glen S. Grepon**
Northern Bukidnon Community College
**PHILIPPINES**
ben.it2c@gmail.com

## ABSTRACT

Computer-based information systems for case management are still at an early stage of adoption in many trial courts in the Philippines. In most cases, information system implemented is the case docket using the official record book on which cases are written and inventory of cases and reports are generated. This is a standalone system that often face data processing, data security and case management challenges. However, there are examples of Information systems in overcoming these pitfalls and producing innovative solutions that surpass data management practices in in many trial courts in the country. One such case is the Regional Trial Court Branch 23 of Cagayan de Oro City in Northern Mindanao, Philippines. A project named Web-based Case Docket Information System (WCDIS) has been designed and developed for the court branch. This system uses a framework known as System Development Life Cycle (SDLC) which is a guide for the design and development. This paper also discusses the key system functionalities and the implementation methodology, including both the benefits and shortcomings of this approach, with the goal of applying lessons learned for future installations. Foremost among the successes of this project is its ability to increase efficiency and reliability in completing court transactions.

## INTRODUCTION

Judicial Courts in the Philippines are at an early stages of development for computer-based case management and case transactions. Information Technology plays a vital role in the judicial system since it is useful on areas of text creation, storage and retrieval, Improvement on access to Law, Recordings of Court Proceedings, Case Management and producing data for Administrative purposes, continuing education and communications (Ntende, 2005). There are actions from the government wherein there are integrations of IT in the courts, one example is the eCourt program, an initiative to increase court efficiency and transparency by reducing court staff's administrative workload and providing lawyers the litigants easy access to case information (American Bar Association, 2013).

In the lower courts such as the RTC Branch 23 and other court branches in the city of Cagayan de Oro eCourt has not been integrated to its operation. The court still uses a book "case docket" that contains summary the basic information about a litigation, bankruptcy, or other type of court case. In an article written by the Inquirer.net of Inquirer Mindanao dated January 31, 2015, "Two persons were killed—a court employee and a security guard—following the five-hour fire that razed the Hall of Justice building of Cagayan de Oro City." According to Fire Insp. Arnold Lagbas, "This was really a fire hazard place because every time we come here, there were papers all over the place, these fed the blaze" (Lagsa, 2015). The court branch is very dependent to its case dockets which halts some of its operations after the catastrophe because it requires massive time to recover and to trace copies of the cases who were burnt by fire.

For small businesses, cloud servers are the preferred method of storing digital data (Rivera, 2019) while for the Philippine government agencies needs to adopt cloud computing as the preferred ICT deployment strategy The shift to cloud technology improve data-sharing among agencies which enables more effective collaboration across institutions (Chin, 2017). It is more practical to use cloud-based technology and repositories of data to avoid possible losses when it comes to unexpected phenomena. In cloud storage data are stored in servers and can be replicated anytime and it requires low cost in management and maintenance, it also offers accessibility since any devices connected to the cloud can access the saved files and information (Carrol, 2018).

This study aims to design and develop a case docket Information System (CDIS) that stores electronic case related records which are recorded in an actual case docket that has been filed at the Court. The system also creates a central access to the repository of data and information for management that can be used for decision- making. automate the case docketing procedure of the court branch and to create an environment to store all docket information in a single repository of data. The system is available 24/7, as long as the device used is connected to the internet. The system provides ways and means for back-up on which the case records entered in the database could be kept in a file anytime depending on policy and protocol for safe keeping and recovery

**LITERATURE REVIEW**

Government as of today is very optimistic in embracing newer technologies to be implemented in all of its branches; each one has its own version of how e-government should be. There's this concept of having all government offices connected, plus having free Internet in public places. As time goes by, the need for technology is becoming a necessity, and that's why the ICT field is on the move to provide solutions to every need in most of the government offices in the Philippines (Lallana, 2002). The government's Information and Communications Technology Office (ICTO) formally unveiled a new e-Government master plan for the next three years, which the office claimed will promote more transparency and efficiency in government transactions and services (Tuazon, 2013).

The government started to plan and soon to implement a nation-wide computer-based system in all government offices due to the virtue of Executive Order No. 265 which is signed by the President Joseph Ejercito Estrada, counter signed by the information systems plan (GISP) as framework and guide for all computerization efforts in the government. The Government Information Systems Plan (GISP), also to be known as "Philippine Government Online" has be realized and to ensure the full and effective implementation of the GISP, all government agencies and instrumentalities, including local government units, aligns their respective computerization projects with the priorities identified in the GISP. The ITECC or the Information Technology and Electronic Commerce Council prepares consolidated annual program of expenditures for government computerization, which becomes part of the President's Annual Expenditure Program for submission to Congress.

After 12 years come a response and improvement from the Government Information Systems Plan (GISP) that was a framework for "Philippine Government Online." Currently, the move of e-Government has been very strong since 2012 as a Master plan has been given an account. The E-Government Master Plan (EGMP) is a blueprint for the integration of ICTs for the whole government. Looking on the local areas of the government that adopt the drive of having a consolidated government electronic system is the Local Government Unit's adaptation of the e-Government which is called the LGU Project. The overall objective of the eLGU project is to revolutionize local governance by making the LGU electronically enabled, thereby improving the delivery of public service, promoting transparency in government transactions and increasing public access to government information services, among others (DOST-ICTS, 2002).

In line with the Supreme Court's adoption of the e-Government, a similar but very specific system is then given consideration: The Supreme Court launched recently in the Regional and Metropolitan Trial Courts of Quezon City a new pilot program called "eCourt" to automate the trial courts. The aim is to speed up the delivery of justice by reducing case processing time, eliminate sources of graft, and improve public access to performance information in the lower courts (Panganiban, 2013).

A similar type of Information system related to case docketing in the court of USA implemented by The Office of State Courts Administrator (OSCA), contracted with Advanced Computer Technologies (ACT) to create the Florida Drug Court Case Management System (FDCCMS) that was initially used for the adult post-adjudicatory expansion program. (Florida Drug Court, 2015). A feature of FDCCMS includes Web Browser Application that uses secure web-based system that restricts access using Microsoft Active Directory (AD) for user authentication. The U.S. Federal Courts Case Management/Electronic Case Files (CM/ECF) service is a very successful court automation system deployed throughout the country that integrated case management, electronic court case records and documents, and the electronic transmission, and service of court records via the Internet (Greenwood, 2012).

**METHODOLOGY**
SDLC is a common methodology for system development in many organizations, featuring several phases that mark the progress of the systems analysis and design effort (Hoffer et al, 2002). This method allows system development to be more efficient by following processes in step-by-step manner to achieve the goals of a highly effective system. The Waterfall approach to systems analysis and design was the first established modern approach to building a system (Royce, 1970).

**Requirement Gathering/ Specification**
On gathering specification in designing the system, the proponent analyzes the organizational reports and documents of the office, including consolidating the documents the old platform produces. The forms gathered are the following: forms for docket entry, attachment report, inventory reports, and monthly reports. A follow up interview to the key informant is one of the alternative ways of gathering information about the office organizational flow, its services and its primary transactions

**Analysis and Design**
Analyzing the requirements is very important to design and develop the over-all functions of the system. A set of Data Flow Diagram (DFD) provides logical model that shows what the system does, not how it does it (Shelly, Cashman, & Rosenblatt, 2009) it allows us to re-design the process that is adopted by the use of the system and an Entity Relational Diagram (ERD) provides an overall view of the system and a blueprint for creating the physical data structures (Hoffer, George, & Valacich,2011) this diagram allows us to design the database structure of the repository of data.

**Development**
The development phase is when all of the actual code is written; in building up the system in web development, Hypertext Mark-up language (HTML) is the foundation in building the entire web page. The main programming language used is Hypertext Preprocessor (PHP), a coding and a server side scripting language that is used in making a fully dynamic web pages, Cascading Style Sheets (CSS) is to be used to create better designs and Java Script (JS) for building user interface components. These elements are handled by Bootstrap, an open-source front-end web framework for a faster web page build up with pre-designed templates that can be can be customized for a professional and commercial look web pages. Apache web server is used to process request via Hyper Text Transfer Protocol (HTTP) to distribute information on the World Wide Web (WWW), for the repository and storage of case data a Database Management System (DBMS) specifically MySQL is used through its (PHPMyAdmin) as user interface for making databases and tables that can be normalized for the CDIS. The development environment runs on the Windows platform.

**User Acceptance Test**
Acceptance Testing is a tool used to test the functional specifications in simulated live operation. The system is to be tested by the key informants of the office. The key informants are the following: The clerk of court, the legal researcher, and the secretary because they are the person who uses the system for their day to day tasks.

**Deployment**
During deployment, the system is delivered and installed for a state of operational readiness. The system is not hosted on a local setup within the court branch of RTC Branch but instead, it is hosted on a web hosting site. To host the web pages, project files are uploaded in its FTP (File Transfer Protocol) server using the File Manager Program (FMP) of the web host after creating and logging-in in one's account. The file manager console allows upload using a single file or by bulk through extraction of file using a (.zip) extension. Once the files are already uploaded those files are ready to be manipulated and edited to have a dynamic effect on the web pages. The file manager program supports file backup and recovery for safekeeping and duplication of files since it is accessible any time and everywhere as long as one has an internet connection. On the repository of data and records the free host has its own DB Manager Program (DBMP) that supports creation of database and importation of (.sql) files which contains electronic records of the case docket.

**System Evaluation**
Evaluation is a process that critically examines a program. It involves collecting and analyzing information about a program's activities, characteristics, and outcomes. Its purpose is to make judgments about a program, to improve its effectiveness, and/or to inform programming decisions (Patton, 1987). A System Usability Scale (SUS) created by John Brooke has been customized to fit for the Case Docket Information System as a tool for evaluation with Likert-type scale as rating scale. System Usability Scale is a 10-item questionnaire that gives an overview of satisfaction with software. It was developed by John Brooke and it is freely available for use, providing acknowledgement is made from the source.

**RESULTS**

Figure 1 displays the login module. The login module allows authorized user for administration and case manipulation and other activities within the system, unauthorized users are not allowed to even view any cases information.

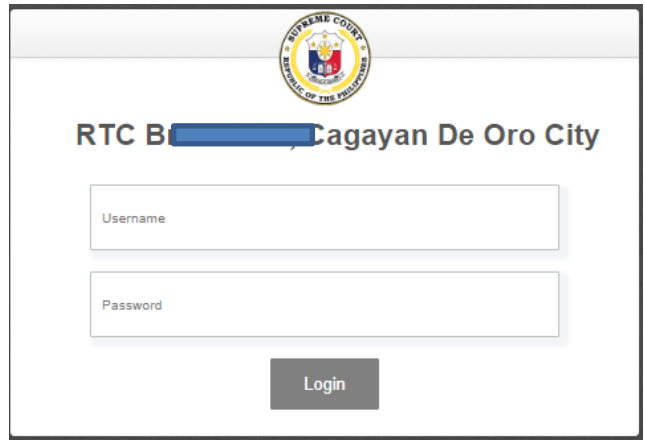

Figure 1. Login Module of the CDIS.

Figure 2 is a screenshot of the administrator dashboard. The dashboard holds all functionalities needed for the system to work. It also aids the user where to go and what to do since controls are on the left tab of the window and activities to do are on the center to right side of the screen. The dashboard holds tool bar items such as Case overview that holds the summary of all cases entered in the database; it also shows the number of case count per section. And generation of reports which is an attachment of their submission to the supreme court.

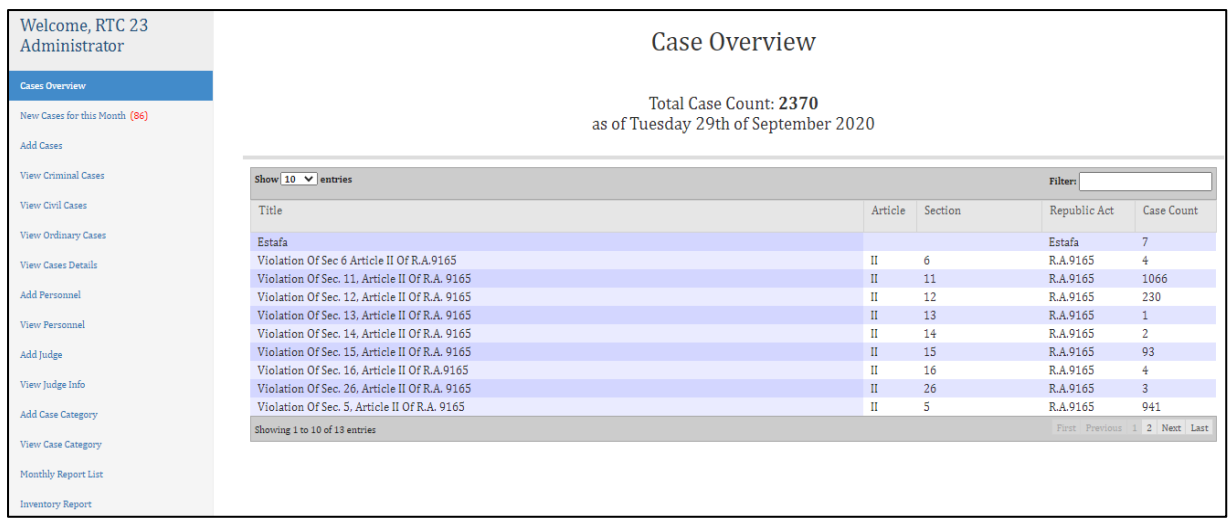

Figure 2. the CDIS Administrator Dashboard.

Figure 3 is a screenshot of the Case Management adds case information both criminal and civil cases, adding of case details, adding of case category, viewing of case records, viewing of case details records and case category, editing and deleting of case records are the functions which is catered in the case management tool bar,

Figure 3. The Case Management Module of the CDIS.

Figure 4 shows the inventory reports used by the court branch as an attachment to their monthly reports which is submitted to the supreme court. This contains the case number, the Nature of the Case, date that the case was filed, dates of arraignment, pre-trial, initial trial and submission for decisions.

Figure 4. Inventory Report generated by the CDIS

**DISCUSSIONS**

Before the User Acceptance Testing, the personnel should know how to use the system. With this, training is conducted. Training provides required skills and knowledge before using the system. In the case of the court branch, a printed system manual is given to the personnel as a reference on what to do or how to work with the system. Then they are allowed to test the system based on the instructions that are written on the system manual.

Table 1 shows the UAT results on areas of administration, the administration performs the following functions: add new personnel, edit personnel, add new judge record, edit judge record, add new case category, edit case category, delete record and search record. See Table 1 for the test case for administration.

Table 1: The Test case for Administration.

| | User-Acceptance Test (UAT) Matrix | | | |
|---|---|---|---|---|
| | Adapted from Department of Infrastructure, Energy and Resources Acceptance Test Plan Matrix | | | |
| SYSTEM NAME: | CASE DOCKET INFORMATION SYSTEM | | | |
| Proponent: | Benzar Glen S. Grepon | | | |
| Test Case: Administration | | Date: October 6, 2016 | | |
| No. | Condition being tested | Expected results | Actual Results | Remarks |
| 1 | Admin> Add new personnel | Add new personnel record | It can add new user | A |
| 2 | Admin > Edit personnel | Edit personnel record | It can edit personnel record as assigned by the administrator | A |
| 3 | Admin> Add new judge | Add new judge record | It can add new judge record | A |
| 4 | Admin> Edit Judge Info | Edit Judge Record | It can edit Judge record as assigned by the administrator | A |
| 5 | Admin> Add new Case Category | Add new Case Category | It can add new Case Category | A |
| 6 | Admin> Edit Case Category | Edit Case Category | It can edit Case Category as assigned by the administrator | A |
| 7 | Delete record | Able to delete a record | It can delete records in cascade | A |
| 8 | Search record | Able to search and filter a record | Both user and administrator can search the record | A |

The files can be downloaded anytime if they need it, in case the actual file is difficult to locate. (See Table 2 for the test case for case entry and file attachment). One unique feature of the system is that it can upload any file for the upload file functionality allows the system to have a document counterpart for the electronic case record which is printed for hardcopies for cabinet filing.

Table 2: The Test Case for Case Entry and File Attachment

| Test Case: Case Entry and File Attachment | | Date: October 6, 2016 | | |
|---|---|---|---|---|
| No. | Condition being tested | Expected results | Actual Results | Remarks |
| 1 | Add Case | Add Case Record | It can add unique case record | A |
| 2 | Add Case Details | Add Case Details | It can add needed details for a case | A |
| 3 | Add Attachment(file) | Attach Actual Softcopy of the case referenced to the case number | It can attach files | A |
| 4 | Edit Case | Edit Case Record | It can edit case record except the case Number | A |
| 5 | Edit Case Details | Edit Monitoring of Status of the case | It can Edit the status of the case | A |
| 6 | Add Additional Attachments | Can add additional File Attachments for file uploading history | It can add additional Attachments and viewed in a list | A |
| 7 | Add File Location of the Attached File (e.g. Cabinet, Box, Folder, etc.) | Can Add File Location for tracking the actual File | Same result | A |
| 8 | Edit File Location of the Attached File | Can Edit the file Location for tracking the actual File | Same result | A |

The last test case is for the Case Search, View Case and Generation of Report as shown in table 3. The following functions allow the user of the system to search specific case based on the search keyword they wish to use for them to filter the exact case for a certain person, and lastly for report generation. Reports in the branch office is very useful since the system can print monthly case reports whether they need criminal case list or civil case list and for the inventory report on which they can print semi-annual record per case.

Table 3: The Test case for Case Search, View Case and Generation of Report with approval from the clerk of court and accepted by the Office Clerk

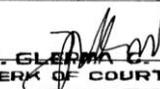

**Evaluation Results**

The evaluation phase of the project is to assure quality of work and usability of the system that matches the expected outcome set by the users during data gathering. Evaluation is conducted after the system is deployed to allow users to test if the system could be used by the office. The test also shows the positive and the negative aspect of the system on its actual performance. Table 3 shows the usability and effectiveness result through System Usability Scale which was developed by John Brooks on 1986.

Table 5 System Usability Scale (SUS) Result

| QUESTION: | User (1) | User (2) | User (3) | User (4) | Average |
|---|---|---|---|---|---|
| 1. I think that I would like to use this system frequently | 5 | 5 | 5 | 5 | 5.00 |
| 2. I found the system very beneficial to the office | 5 | 5 | 5 | 5 | 5.00 |
| 3. I thought the system was easy to use | 5 | 5 | 4 | 5 | 4.75 |
| 4. I think that I would not need the support of a technical person to be able to use this system. | 4 | 4 | 2 | 4 | 3.5 |
| 5. I found the various functions in this system were well united | 5 | 5 | 4 | 5 | 4.75 |
| 6. I thought I can work with less mistake with this system | 5 | 5 | 4 | 5 | 4.75 |
| 7. I would imagine that most people would learn to use this system very quickly | 5 | 5 | 4 | 5 | 4.75 |
| 8. I found the system very Useful | 5 | 5 | 5 | 5 | 5.00 |
| 9. I felt very confident I using the system. | 5 | 5 | 5 | 5 | 5.00 |
| 10. I don't need to learn a lot of things before I could get going with this system | 5 | 5 | 3 | 5 | 4.5 |
| | | | | **GENERAL AVERAGE:** | **4.70** |

Table 5 shows the actual scores of the System Usability Scale (SUS). The data on the evaluation, that was conducted and answered by the four (4) participants of the court branch, were collated and summarized into a single table.

The researcher used the arithmetic mean method to be able to come up with the result. To be able to get the average, the researcher adds up all the numbers and then divide it by the number of the numbers or how many numbers there are.

**Regular use of the system**
On question 1 of frequent usage, the respondents agreed to score a perfect 5 simply because every week filing of cases is continually performed in the case docket. The personnel handling case filing needs to use the system to input the newly filed cases to update the repository of case records for easy management.

**Independency of the personnel that uses the system**
On questions number 4, regarding independency of using the system the majority did score perfect and one scored 2 because most of the personnel in the court is not that confident in using a computer without any aid from someone who has broad knowledge in computer usage and troubleshooting in case of irregularities.

**Beneficial Contribution of the system in the workplace**
Majority of the respondents say that the system is very beneficial in the workplace based on the results in questions number 2 and 8 which are rated 5.0 since it provides fast and reliable results especially in generating of case reports, and also the system can monitor the status of the cases, and they can give feedbacks through the system to those who want to know the updates and movements of the case.

**User-friendliness of the system**
On questions 3,5,7,9 and 10 the personnel think that the system is easy to use by an average of 4.5 because they are exposed on how to navigate and surf the web. The template used is a dashboard type which is common to most websites especially Facebook on which they are very fond of. Respondents also say they are confident in working with the system since they are given manuals on which they can easily follow the steps provided effectively, and lastly because they are already familiar with all functionalities since they are given time to witness the demonstration during the training conducted, and they are given chance to do a hands-on on how to use the system.

**Over-all impression of the system**
The personnel were so grateful that they were given the opportunities to be the first court branch in the Cagayan de Oro to use the CDIS since majority of the court branch in Cagayan de Oro City is still dependent on its case docket for case monitoring and generation of reports. The clerk of court also recommended the system to other court branch because aside from making their work faster and easier, it also helps them to backup necessary files related to the case record that if any incident the same as what happened in the year 2015 when a fire razed throughout the Hall of Justice building. They are not in fear that their case docket together with the supporting documents might be lost and be burned up again, and they can't do anything about it. All their files and records can be hosted already in the internet which gives them security that they can restore anything even if another incident hits the working place.

In summary, it is quite evident that the result provides a clear understanding of the real effect of the system in the workplace, looking at general average computed which reaches 4.70 computed using the basic averaging. It scores very good, which means the system provides most of what the whole court needed.

**CONCLUSIONS**

Automating the judicial processes is a big task and very challenging but making it into a reality is a thing to look forward to since it makes the work in the judicial system especially in the Philippines faster and more efficient. CDIS is one of the small chunks in making it possible since it operates in a smaller court branch which works on a specific type of cases.

The CDIS plays an important role in the court's current process since it helps in making court case docket book records into a computer-based system which is easier to use and very accessible because it only requires an internet connection and a device capable of connecting to the internet. Security wise, the system can be managed and operated by the office personnel with login authentication. Test cases are performed prior to deployment to ensure user acceptance to see if the system fits their processes and their office needs. Training of the users was performed after the test cases and turn-over of the system manual was conducted after the training. Transactions are real-time since files and folders together with the database are hosted online through the web hosts with its control panel which can be accessed anytime through the World Wide Web. Evaluation was conducted after the deployment to ensure usability of the system, looking through with an average above 4.7 as the summarized result from the 4 personnel who evaluated the system. This means the system achieved its purpose and is usable.

The strength and success of the information system rely on the users' satisfaction. The cooperation and understanding of the system users is a big factor for the success of the system. Failure happens if there is no proper implementation conducted in the organization. The goal of this study is to present what information system can contribute in order to help and solve the problem of the court branch.

**ACKNOWLEDGEMENT**

"All wisdom comes from the Lord, and so do common sense and understanding. God gives helpful advice to everyone who obeys him and protects all of those who live as they should. God sees that justice is done, and he watches over everyone who is faithful to him." (Proverbs 2: 6-8). I Thank GOD for all wisdom, knowledge and strength. I also thank the people who encouraged me and supported me in this research:Cyril Jane C. Ranido, MIT, MIM, Dean of the College of Computer Studies, Capitol University for the moral support and encouragement to continue my graduate studies. Engr. Gerardo S. Doroja, Dean of College of Computer Studies, Xavier University and Sir Joseph Anthony C. Sabal my mentor, who were my (Torch) that lights and guides me to understand the real value of Research in the IT field; The RTC Branch 23 Personnel for entertaining my requests for my data gathering especially to Ms. Patsy Pantanosas who contributed a lot in giving me all possible information and knowledge about the office and its transactions. My Colleagues in the College of Computer Studies of Capitol University and Northern Bukidnon Community College for both professional and civil support showed on my efforts towards this project. My parents, siblings and to my wife for always there to support and love.